\documentclass[%
reprint,
 amsmath,amssymb,
 aps,
]{revtex4-2}
\usepackage{color}
\usepackage{graphicx}% Include figure files
\usepackage{dcolumn}% Align table columns on decimal point
\usepackage{bm}% bold math
\usepackage[colorlinks,
            linkcolor=blue,
            anchorcolor=blue,
            citecolor=blue,
            urlcolor=blue]{hyperref}% add hypertext capabilities
\usepackage[percent]{overpic}
\begin{document}

\preprint{APS/123-QED}

\title{Leapfrogging \textit{Sycamore}: Harnessing 1432 GPUs for 7$\times$ Faster Quantum Random Circuit Sampling}% Force line breaks with \\
%\thanks{A footnote to the article title}%give a star to title

\author{Xian-He Zhao$^{1,2,3,4}$}\altaffiliation{Equally contributed to this work.}
\author{Han-Sen Zhong$^{4}$}\altaffiliation{Equally contributed to this work}\email{zhonghansen@pjlab.org.cn}
\author{Feng Pan$^{2}$}\altaffiliation{Equally contributed to this work.}
\author{
Zi-Han Chen\textsuperscript{1,2,3}, 
Rong Fu$^4$,
Zhongling Su$^4$,
Xiaotong Xie$^4$,
Chaoxing Zhao$^4$,
Pan Zhang$^5$,
Wanli Ouyang$^4$,
Chao-Yang Lu\textsuperscript{1,2,3},
Jian-Wei Pan\textsuperscript{1,2,3}
}
\author{Ming-Cheng Chen\textsuperscript{1,2,3,}}\email{cmc@ustc.edu.cn}

\affiliation{
$^1$Hefei National Research Center for Physical Sciences at the Microscale and School of Physical Sciences,
University of Science and Technology of China, Hefei 230026, China\\ 
$^2$Shanghai Research Center for Quantum Science and CAS Center for Excellence in Quantum Information and Quantum Physics,
University of Science and Technology of China, Shanghai 201315, China\\
$^3$Hefei National Laboratory, University of Science and Technology of China, Hefei 230088, China\\
$^4$Shanghai Artificial Intelligence Laboratory, Shanghai, 200232, China\\
$^5${CAS Key Laboratory for Theoretical Physics}, {Institute of Theoretical Physics, Chinese Academy of Sciences}, {{Beijing}, {100190}, {China}}\\
}

\date{\today}% It is always \today, today,

\begin{abstract}
Random quantum circuit sampling serves as a benchmark to demonstrate quantum computational advantage. Recent progress in classical algorithms, especially those based on tensor network methods, has significantly reduced the classical simulation time and challenged the claim of the first-generation quantum advantage experiments. However, in terms of generating uncorrelated samples, time-to-solution, and energy consumption, previous classical simulation experiments still underperform the \textit{Sycamore} processor. Here we report an energy-efficient classical simulation algorithm, using 1432 GPUs to simulate quantum random circuit sampling which generates uncorrelated samples with higher linear cross entropy score and is 7 times faster than \textit{Sycamore} 53 qubits experiment. We propose a post-processing algorithm to reduce the overall complexity, and integrated state-of-the-art high-performance general-purpose GPU to achieve two orders of lower energy consumption compared to previous works. Our work provides the first unambiguous experimental evidence to refute \textit{Sycamore}'s claim of quantum advantage, and redefines the boundary of quantum computational advantage using random circuit sampling. 

\end{abstract}

\maketitle

%\tableofcontents

\section{INTRODUCTION}\label{sec1}

Quantum computers represent a new paradigm of computing, and in theory promise to solve certain problems much faster than classical computers ~\cite{feynman1985quantum, Shor94}. A major milestone is to discover quantum algorithms for noisy intermediate-scale quantum computers and demonstrate the long-anticipated quantum computational advantage (QCA)~\cite{Preskill2018quantumcomputingin}. Such algorithms include boson sampling and its variant~\cite{aaronson2011computational, hamilton2017gaussian, quesada2018gaussian}, random circuit sampling (RCS)~\cite{boixo_characterizing_2018}, and instantaneous quantum polynomial sampling~\cite{bremner2016average}. Using superconducting circuits ~\cite{Arute2019, morvan2023phase,Zuchongzhi2.0_PhysRevLett.127.180501, zuchongzhi2.12022240} and photons ~\cite{jiuzhang1.0, jiuzhang2.0, jiuzhang3.0, Borealis2022}, increasingly sophisticated experiments have provided strong evidence of QCA. For example, in Google’s 2019 groundbreaking experiment, the \textit{Sycamore} processor obtained 1 (3) million uncorrelated samples in 200 (600) seconds, with a linear cross entropy (XEB) of 0.2\%. It was then estimated that simulating the same process on \textit{Summit} supercomputer would cost 10,000 years \cite{Arute2019}. XEB estimates circuit fidelity by comparing the sample distribution from experiments with theoretical predictions. We elaborate upon XEB, and the post-processing step to increase it, in the subsequent METHODS section.

\begin{figure}[htb]
\centering
\includegraphics[width=0.45\textwidth]{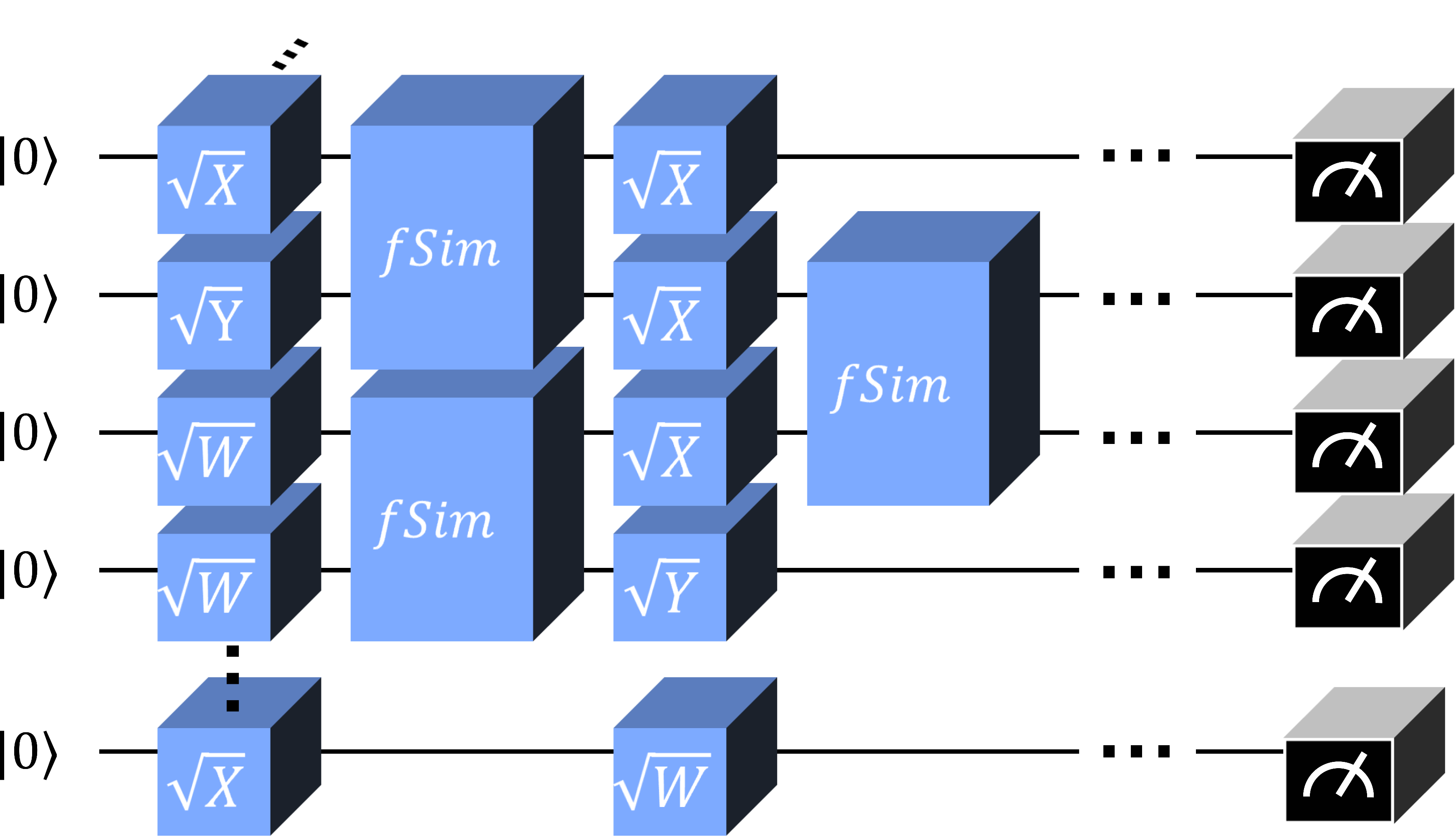}% Here is how to import EPS art
\caption{\label{circuit} A schematic diagram of a quantum random circuit. The single-qubit gates are randomly select from $\{\sqrt{X}, \sqrt{Y}, \sqrt{W}\}$.
}
\end{figure}

Similar to the Bell experiments, the QCA is not a single-shot achievement but expects continued competition between improved classical simulation algorithms and upgraded quantum hardware. During this process, the QCA milestone can be progressively better established. For the random circuit sampling, shown as Fig. \ref{circuit}, emerging classical algorithms~\cite{Liu10.1145, Pan2022PhysRevLett.129.090502, PhysRevLett.128.030501, PRXQuantum.5.010334, oh2023tensor, huawei2022multitensor} based on tensor networks have significantly reduced the time-to-solution and mitigated the exponential growth of memory demands, and have challenged the claim of first-generation of quantum advantage. 
These algorithms leverage the low XEB value (0.002) of the \textit{Sycamore} experiment to employ low-fidelity simulations, emerging as an effective strategy for reducing complexity.

\begin{figure}[tb]
\centering
\includegraphics[width=0.45\textwidth]{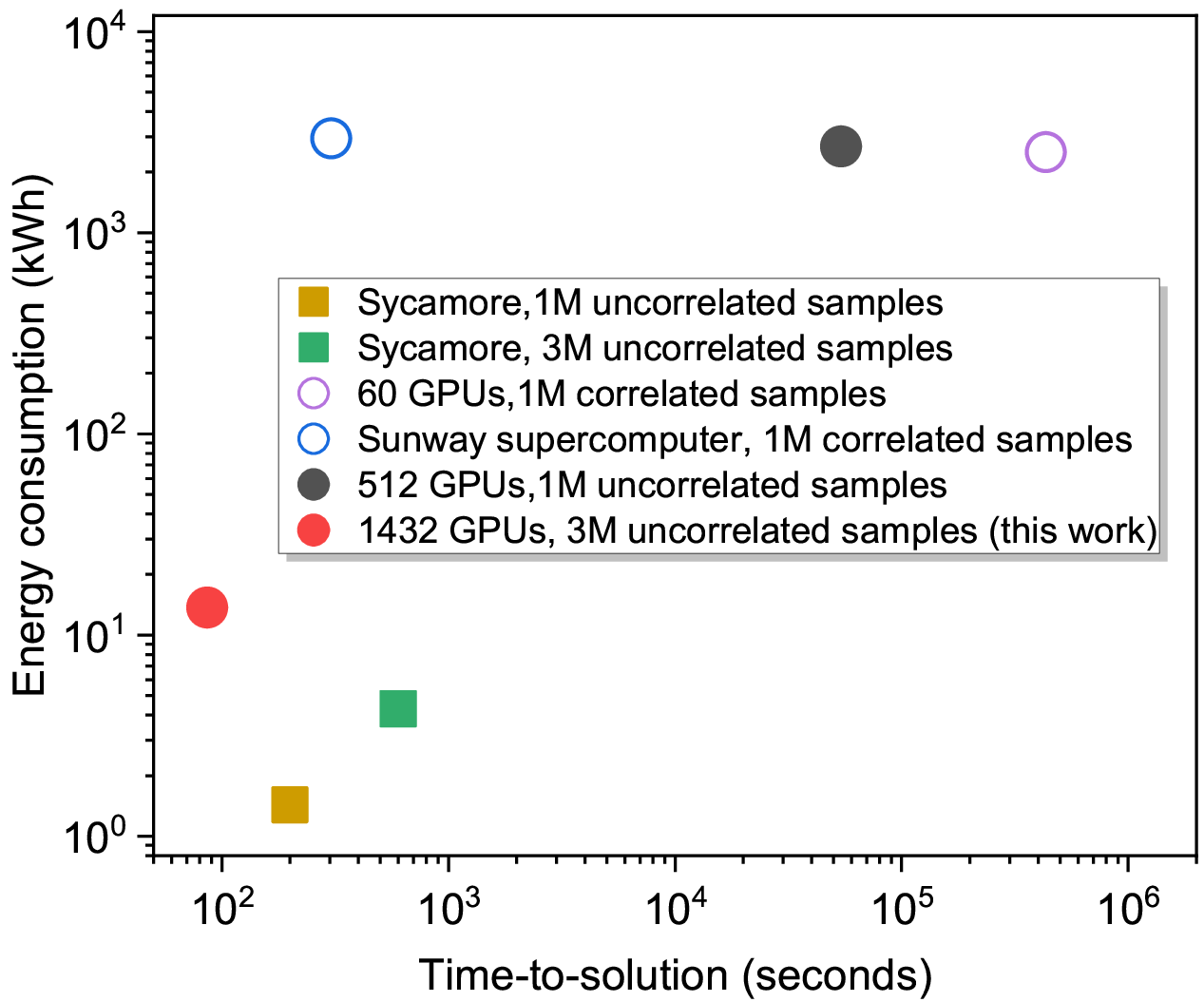}% Here is how to import EPS art
\caption{\label{comparison} Performance of implementations of sampling the \textit{Sycamore} circuit. The horizontal axis donates the time-to-solution, and the vertical axis donates the energy consumption in the quantum experiment or classical simulations. Circles and squares correspond to classical simulations and quantum experiments, respectively. 
The hollow circle indicates a correlated sampling loophole 
in the corresponding classical simulation.
}
\end{figure}

Among the previous attempts, summarized in Fig. \ref{comparison}, Yong \textit{et al.} \cite{Liu10.1145} used the \textit{Sunway} supercomputer to perform the classical simulation, and obtained 1 million samples in 304 seconds -- still slower than \textit{Sycamore}. More importantly, their obtained samples were \textit{correlated}, which was not equivalent to a true quantum experiment. Xun \textit{et al.} \cite{PRXQuantum.5.010334} completed the simulation in 0.6 seconds, by simplifying the circuit into two independent parts, which, however, resulted in a low XEB value of $1.85\times 10^{-4}$, failing to meet the benchmark set by \textit{Sycamore}. Pan \textit{et al.} \cite{Pan2022PhysRevLett.129.090502} employed 512 GPUs to successfully obtain 1 million \textit{uncorrelated} samples \cite{zlokapa_boundaries_2023} with an XEB value of 0.0037, but at a time cost of 15 hours.

In addition to the time-to-solution, the XEB, and the sample correlation, another factor worth mentioning is energy consumption \cite{PRXQuantum.3.020101}. The energy consumption of previous work \cite{Liu10.1145, Pan2022PhysRevLett.129.090502, PhysRevLett.128.030501, PRXQuantum.5.010334, oh2023tensor, huawei2022multitensor} is approximately three orders of magnitude higher than the quantum processor. For practical reasons, reducing the power overhead is also of great interest.

\section{SUMMARY OF RESULTS}\label{sec2}

In this work, we propose a new algorithm that obtains
% loophole-free 
samples with XEB values comparable to that of \textit{Sycamore} and achieves better performance in running time and energy cost, thereby experimentally refuting Google’s quantum advantage claim \cite{Arute2019} in both the solution time and the energy cost. Compared to previous high XEB uncorrelated sampling works, we reduce the energy consumption by two orders of magnitude. Our algorithm uses partial network contraction to achieve low-complexity approximation \cite{pan2023efficient}. According to the insight of Porter-Thomas distribution and a flaw in the XEB measure, we develop a post-processing algorithm to increase the XEB values. We also notice some works mentioned the similar method~\cite{PhysRevLett.128.030501, PRXQuantum.5.010334, tanggara2024classically}. We find that the computational complexity of the optimal contraction scheme is approximately inversely proportional to the storage space, and develop an advanced 8-GPU parallel tensor contraction algorithm that increases the accessible storage space to $8 \times 80~\mathrm{GB} =640 ~\mathrm{GB}$ to reduce the computational complexity without significantly increasing the communication time. Finally, using 1,432 NVIDIA A100 GPUs, we obtain 3 million uncorrelated samples with XEB values of $2\times 10^{-3}$ in 86.4 seconds, consuming only 13.7 kWh of electricity in total. In contrast, \textit{Sycamore} performs the same task to obtain 3 (1) million samples in 600 (200) seconds, consuming 4.3 kWh of electricity for cooling water~\cite{Arute2019}.

{METHODS}\label{sec3}

\begin{figure}[htb]
\centering
\includegraphics[width=0.45\textwidth]{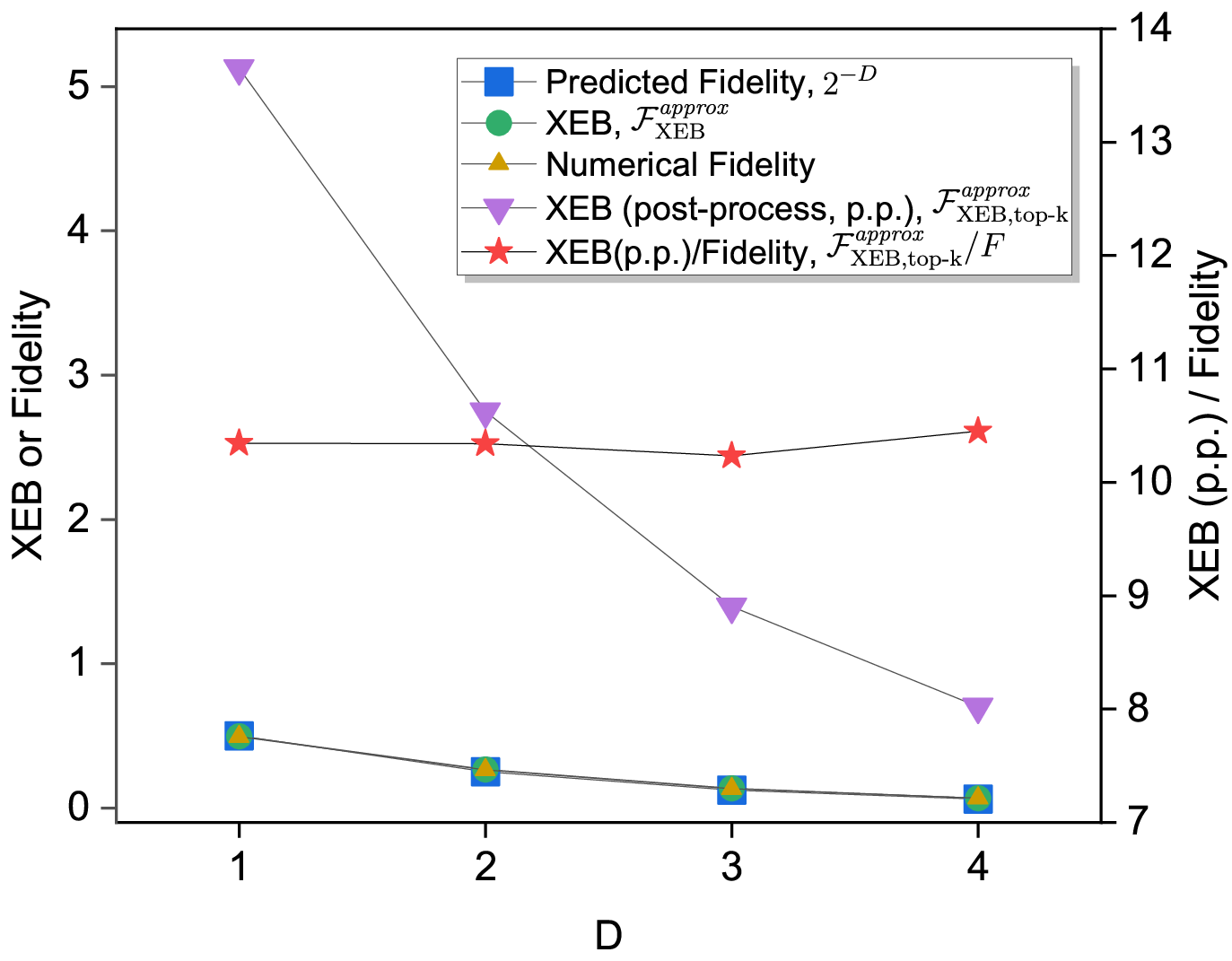}
    % \begin{overpic}[width=0.44\linewidth]{pic/correlation(2^-1).png}
    %     \put(0,95){\large\textbf{a}}
    % \end{overpic}
    % \hspace{2mm}
    %     \begin{overpic}[width=0.46\linewidth]{pic/correlation(256^-1).png}
    %     \put(0,91){\large\textbf{b}}
    % \end{overpic}
    % \hspace{2mm}
    %     \begin{overpic}[width=1\linewidth]{pic/Post_process_effect.png}
    %     % \put(0,63){\large\textbf{c}}
    %     \label{correlation of approxi and ideal}
    % \end{overpic}
    % \hspace{2mm}
\caption{\label{correlation}
% (a) and (b) illustrate the correlation between the real part of the exact amplitudes and the approximate amplitudes for breaking 1 edge (a) and breaking 8 edges (b). The correlation observed for the imaginary part is analogous to that of the real part. (c) 
Simulation fidelity (theoretically predicted as well as numerically calculated) and the XEB values of samples obtained via the MCMC sampling method and the top-k method respectively under the circumstance of contracting $2^{-D}$ of sub-networks for various positive integers $D$. The circuit under consideration consists of 30 qubits with 14 layers of gate operations. The numerical fidelity is computed using approximate amplitudes and exact amplitudes. 
$\mathcal{F}_{XEB}^{approx}$ is the XEB value of the $2^{20}$ samples obtained from $2^{20}\times 2^{6}$ bitstrings via MCMC sampling and $\mathcal{F}_{XEB,top-k}^{approx}$ is the XEB value of the $2^{15}$ samples selected from $2^{30}$ bitstrings via the top-k method. 
% XEB is calculated based on $2^{20}$ samples from a total of $2^{20} \times 2^6$ samples, while XEB(p.p.) is calculated by using the top $2^{15}$ samples from $2^{30}$ samples. 
The post-processing (p.p.) step increases XEB, enabling XEB to exceed 1 for circuits with a small D value. }
\end{figure}

\subsection{Composition of our algorithm}

Our algorithm works in the following way. First, we calculate the approximate output probability $p(s)$ of $k$ bitstring groups by contracting a fraction $f$ of the tensor network, where each group includes $2^l$ correlative bitstrings. These preprocessed samples are generated by iterating over all $l$-bit strings for $l$ qubits, which are called open qubits, and sampling uniformly over the rest of qubits for $k$ times. We obtain a sample sequence $\tilde{\mathcal{S}} = \{s_{L}^{i}|L\in\{1,\cdots,k\},i\in\{1,\cdots,2^l\}\}$. Then, based on the top-$k$ method, we obtain the desired sequence of samples $\mathcal{S}$ by selecting the elements in $\tilde{\mathcal{S}}$ with the top $k$ largest ${p}(s)$ values. Finally, we obtain $k$ uncorrelated bitstrings, whose XEB is $f\cdot \ln (|\tilde{\mathcal{S}}|/k)$ (see Eq.~\ref{eq: XEB top-k}). We notice that there are some works that also use the similar method to emulate Sycamore~\cite{PhysRevLett.128.030501, PRXQuantum.5.010334, tanggara2024classically}. In our work, we will give a more systematic proof both theoretically and numerically.

The tensor network is divided into many sub-networks via the slicing method~\cite{huang_efficient_2021,pednault2017pareto, chen2018classical, slice_overhead}, which is to fix the indices over certain edges for each sub-network such that each sub-network can be contracted independently and the space complexity for each contraction is reduced while increasing the total time complexity. 
We only contract a fraction of the sub-networks. The approximate simulation is similar to the idea in previous work~\cite{Pan2022PhysRevLett.129.090502}. This method significantly reduces the computational complexity at the expense of a decrease in fidelity. 

 \begin{figure}[htb]
\centering
\includegraphics[width=0.45\textwidth]{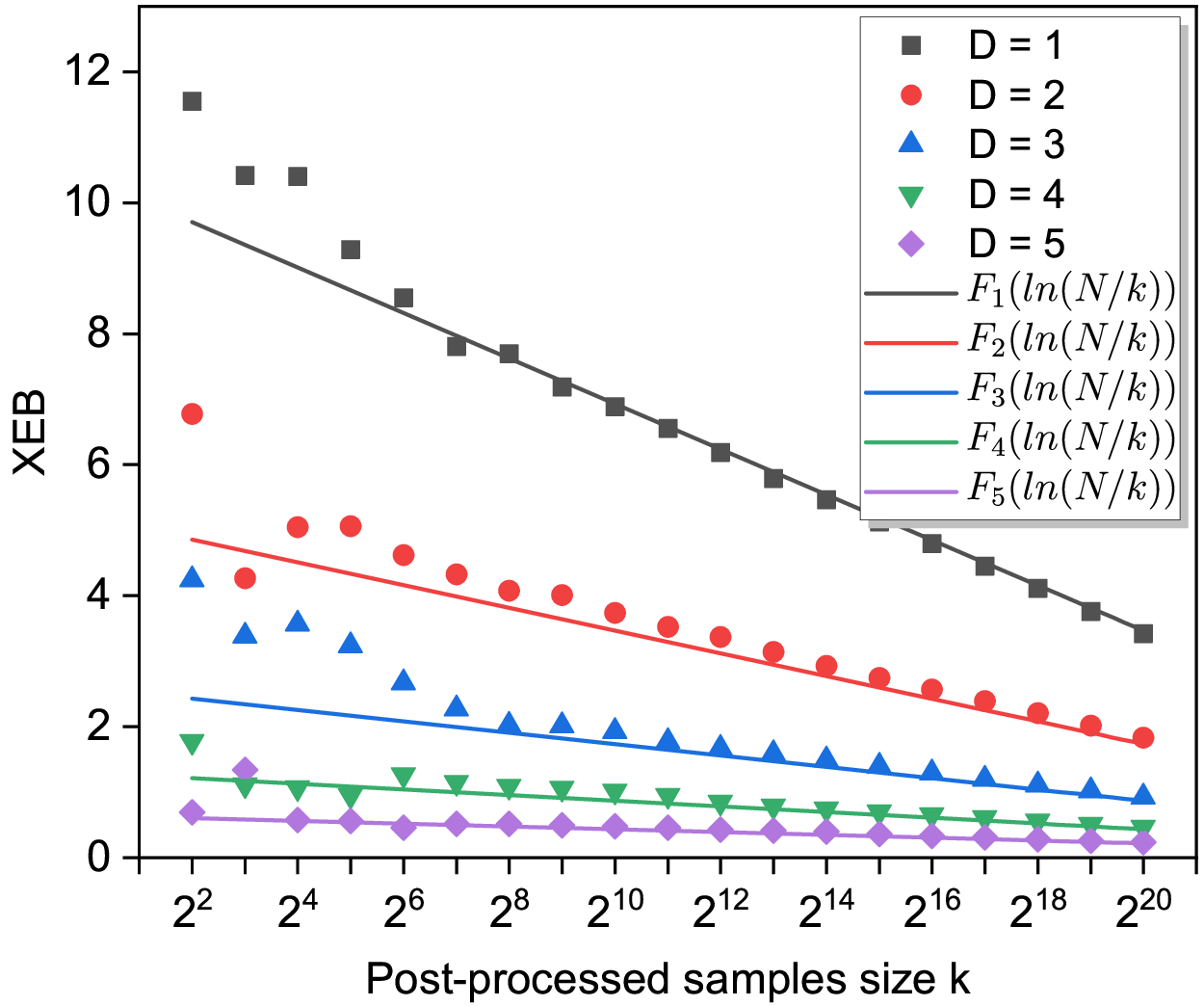}% Here is how to import EPS art
\caption{\label{theoretical_vs_numerical_XEB} The theoretical and numerical results of XEB(p.p.) are presented for different sub-network contraction fractions $f=2^{-D}$ of sub-network, where the number of $D$ increases from 1 to 5. The theoretical results are calculated using Equation \ref{eq: XEB top-k}, where $F_D \approx 2^{-D}$ represents the fidelity, $N=|\tilde{\mathcal{S}}|=2^{30}$ is the sequence size.}
\end{figure}

Notably, as is discussed in detail later, the top-$k$ method amplifies the XEB value by distilling from $\tilde{\mathcal{S}}$ and, as a result, significantly alleviates the requirement on the fidelity of the classical simulation.

\subsection{Post-processing method and XEB amplification}

In the quantum RCS problem, the fidelity of a sequence of samples $\mathcal{S}$ is estimated by the XEB~\cite{boixo_characterizing_2018}, and the definition of XEB ($ \mathcal{F}_{\mathcal{XEB}}$) is:
\begin{equation}\label{eq: XEB}
    \mathcal{F}_{\mathcal{XEB}}(\mathcal{S}) := \frac{N}{|\mathcal{S}|}\sum_{s\in\mathcal{S}}q(s) - 1,
\end{equation}
where, 
$N=2^n$ is the number of all possible bitstring outcomes with n being the number of qubits and $q(s)$ is the ideal output probability for bitstring $s$. The ideal output probability $q$ is predicted to be Porter-Thomas (PT) probability distribution~\cite{boixo_characterizing_2018} given by $P(Nq) = e^{-Nq}$. If a large enough number of samples are obtained according to the ideal output probability, then their XEB value should approach $1$~\cite{Arute2019}. In our algorithm, in an ideal setting where we calculate the probability distribution for each string in $\tilde{\mathcal{S}}$ accurately, $\mathcal{S}$ is obtained as the top $k$ strings in $\tilde{\mathcal{S}}$ with the largest output probability. Using the fact that the distribution of the output probabilities approaches the PT distribution for deep circuits, we prove in Appendix~\ref{appendix: The theory value of XEB
after post-processing} that 
\begin{equation}
\label{eq: top-k ideal}
    \langle\mathcal{F}_{\text{XEB,top-k}}^{\textit{ideal}}(\mathcal{S})\rangle = \ln (|\tilde{\mathcal{S}}|/|\mathcal{S}|)= \ln (|\tilde{\mathcal{S}}|/k).
\end{equation}
Hence, in the ideal case, the top-k method amplifies the XEB value by about $\ln (|\tilde{\mathcal{S}}|/k)$. Moreover, for noisy quantum random circuits, the XEB value is argued to be a good approximation to the circuit fidelity~\cite{Arute2019}. 
% Since each sub-network contributes equally to the fidelity. Consequently, the total fidelity is determined by the number of contraction sub-networks ~\cite{Pan2022PhysRevLett.129.090502}. Thus, we expect our classical approximate simulation that achieve fidelity $F$ by summing over a fraction $f$ of the total tensor network, we would obtain samples with an XEB value of 
As for classical simulations, since each sub-network contributes equally to the final fidelity~\cite{markov2018quantum}, we can achieve fidelity $F=f$ by summing over a fraction $f$ of all sub-networks. In this way, we expect approximate simulation would obtain samples with an XEB value of
\begin{equation}
    \label{eq: XEB as an estimate of F for direct sampling}
    \langle\mathcal{F}_{\text{XEB}}^{\textit{approx}}(\mathcal{S})\rangle= F= f.
\end{equation}
via Markov chain Monte Carlo (MCMC) sampling~\cite{Pan2022PhysRevLett.129.090502}. Based on Eq.~\ref{eq: top-k ideal}, we also expect that our algorithm, which combines  approximate simulation with the top-k method, can achieve an XEB value of
\begin{equation}
    \label{eq: XEB top-k}
    \langle\mathcal{F}^{\textit{approx}}_{\text{XEB,top-k}}(\mathcal{S})\rangle = F\cdot \ln (|\tilde{\mathcal{S}}|/k),
\end{equation}
where $F$ is, again, the fidelity of the classical simulation. 
% This could be understood as that the probability of picking out the samples with top $k$ $q(s)$ according to the $p(s)$ of samples is F. 
The factor $F$ in Eq.~\ref{eq: XEB top-k} can be heuristically understood as the probability of the post-processed samples according to p(s) belonging to the sequence with the top $k$ q(s). 
By comparing $\langle\mathcal{F}^{\textit{approx}}_{\text{XEB,top-k}}(\mathcal{S})\rangle$ to $\langle\mathcal{F}_{\text{XEB}}^{\textit{approx}}(\mathcal{S})\rangle$, we observe the amplification factor provided by the top-k method on the XEB value is approximately $\ln(|\tilde{\mathcal{S}}|/k)$. Thus, to reach a certain XEB value, the top-k method reduces the required simulation fidelity by a factor of $[\ln(|\tilde{\mathcal{S}}|/k)]^{-1}$, which directly translates to a reduction of computational cost by the same factor. 

To validate our expectations, we perform numerical experiments on a small-scale random circuit, structurally similar to the \textit{Sycamore} circuit, on 30 qubits with 14 layers of gates. In all numerical experiments, we use the approximate simulation method described in the previous section to evaluate output probabilities of strings in $\tilde{\mathcal{S}}$. First, for MCMC sampling, we create $\tilde{\mathcal{S}}$ with $|\tilde{\mathcal{S}}|=2^{20}\times 2^{6}$ and obtain $2^{20}$ samples for various degrees of approximation controlled by contracting different fractions of sub-network. We observe, in accordance with Eq.~\ref{eq: XEB as an estimate of F for direct sampling}, that for each fraction $f=2^{-D}$ with $D\in\{1,2,3,4\}$, the XEB value $\mathcal{F}_{\text{XEB}}^{\textit{approx}}$, numerical fidelity and analytically predicted fidelity $2^{-D}$ all agree well with each other (Fig.~\ref{correlation}). As for sampling via the top-$k$ method, we simply select $k=2^{15}$ strings with the top $k$ greatest $p(x)$ values from $2^{30}$ strings as the samples. We observe that $\mathcal{F}^{\textit{approx}}_{\text{XEB,top-k}}/F$ centers around the same value for all $D$ (Fig.~\ref{correlation}) and calculate the numerical value of $\mathcal{F}^{\textit{approx}}_{\text{XEB,top-k}}/F$ to be $10.4694 \pm 0.2635$ which is close to the value $\ln 2^{15}=10.3972$ predicted by Eq.~\ref{eq: XEB top-k}. We also examine the dependence of $\mathcal{F}^{\textit{approx}}_{\text{XEB,top-k}}$ on $k$ and observe that for $k\geq 2^{8}$, the numerical data of $\mathcal{F}^{\textit{approx}}_{\text{XEB,top-k}}$ fit extremely well with the lines predicted by Eq.~\ref{eq: XEB top-k} for various $D$ (Fig.~\ref{theoretical_vs_numerical_XEB}). We expect that the deviation for smaller $k$ from Eq.~\ref{eq: XEB top-k} is due to statistical fluctuations more eminent at small sample sizes and that for larger sample sizes, i.e., larger $k$, Eq.~\ref{eq: XEB top-k} should hold well. 

 \begin{figure}[htb]
\centering
\includegraphics[width=0.45\textwidth]{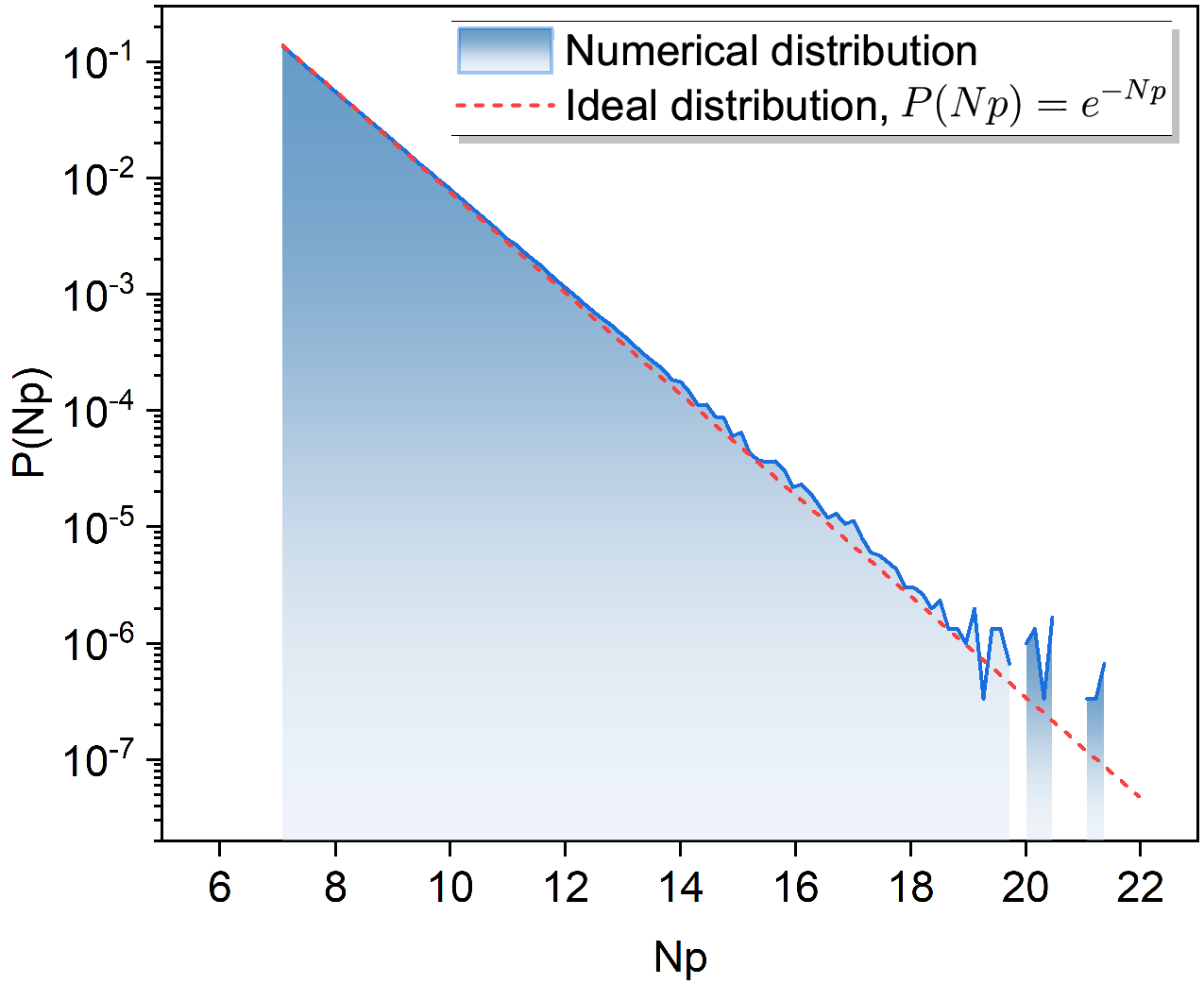}% Here is how to import EPS art
\caption{\label{distribution} Histogram of the probabilities of 3 million post-processed samples from the \textit{Sycamore} circuit with 53 qubits and 20 cycles. The probabilities $|\varphi(s)|^2/f$ are normalized according to the fraction of tensor network to be contracted. The red dotted line denotes the Porter-Thomas distribution.}
\end{figure}

Fig.~\ref{distribution} is the histogram of normalized circuit-output probabilities $|\varphi(s)|^2/f$, corresponding to the bitstrings in the post-processed samples $\mathcal{S}$ and $\varphi(s)$ is the amplitude of bitstring $s$, which is observed to fit PT distribution well. 
This is the result of the following two observations. First, the values of the approximate output probabilities $p(s)$ also satisfy the PT distribution~\cite{Pan2022PhysRevLett.129.090502}. Secondly, the top $k$ values selected from probabilities with the PT distribution still satisfy the PT distribution.

\subsection{Optimizing tensor contraction}

Contraction of complex tensor networks is a difficult task, as different contraction pathways require significantly different computational resources. When simulating quantum advantage experiments using tensor networks, the primary challenge lies in finding the most computationally efficient contraction path within the constraints of limited memory. Pioneering efforts have made substantial progress, such as simplifying tensor networks based on outcome bitstrings and optimizing contraction paths through simulated annealing searches. 

\begin{figure}[htb]
\centering
\includegraphics[width=0.45\textwidth]{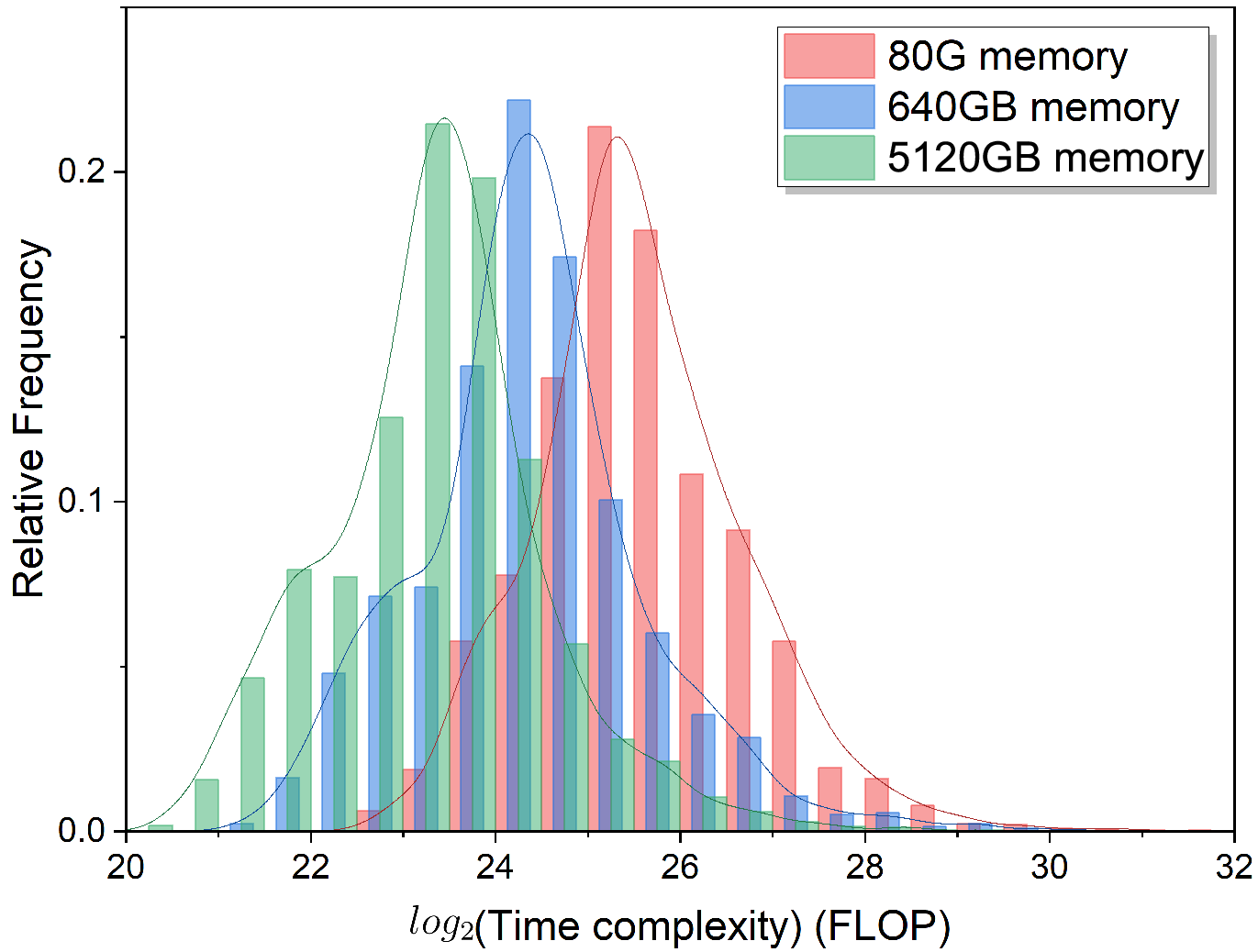}% Here is how to import EPS art
\caption{\label{Time_complexity_with_memory} 
The time complexity distribution of contraction paths is examined under memory constraints of 80 GB, 640 GB, and 5120 GB. Notably, as the available memory increases by a factor of 8, the time complexity of the optimal contraction path decreases by half. This observation highlights the impact of memory resources on the efficiency of contraction operations.
}
\end{figure}

The slicing technique~\cite{huang_efficient_2021,pednault2017pareto, chen2018classical, slice_overhead} allows for the reduction of memory consumption during contractions to fit within available device memory. Tensor network slicing involves selectively calculating only one of the two possible values for certain edges within the network during contractions, and then summing all possible value combinations at the end. For example, when removing edges $i$ and $j$ from tensor network TN, we have 
\[TN = TN_{i=0,j=0} + TN_{i=0,j=1} + TN_{i=1,j=0} + TN_{i=1,j=1} .\]
Slicing reduces the number of indices during tensor contraction, lowering memory consumption. However, sliced edges are the last to be contracted, restricting the flexibility of contraction pathways and resulting in increased computational resources. This is essentially a trade-off between time and space. Numerical experiments show that under memory constraints of 80 GB, 640 GB and 5,120 GB, the computational time complexity of optimal contraction path is approximately inversely proportional to the maximum memory size, shown in Fig~\ref{Time_complexity_with_memory}. Tensor contractions are fetch-intensive tasks. We use 80GB A100 GPUs with an intra-GPU memory bandwidth of 2 TB/s. Eight A100 GPUs within each node are interconnected via NVLink with an intra-GPU speed of 600 GB/s, while nodes are connected via InfiniBand with an intra-node speed of 25 GB/s. We adopt a node-level computation approach, using $8\times80 \text{GB} = 640 \text{GB}$ of memory for contraction computations to achieve a balance between calculation and communication.

We visualize the optimal contraction tree, which consists of a stem path leading from a leaf node to the root node. Nodes on the stem path consume significantly more computation and memory resources than other nodes. To reduce inter-GPU communication, we distributed tensors of the stem nodes across eight GPUs based on the first three dimensions. During a single tensor contraction, if the first three dimensions are not contracted, no inter-GPU communication is required, and each of the eight GPUs computes its respective part. If some of the first three dimensions are contracted, we permute the tensor, swapping the first three dimensions with the next three, and distribute the tensor accordingly with the new three dimensions. To minimize communication overhead, we optimize the order of dimensions within the stem path, ensuring that the three dimensions used for distributed storage persist through as many steps as possible without undergoing contraction. In the final stages of the stem path, we follow the method by Pan \textit{et al.}~\cite{Pan2022PhysRevLett.129.090502}, calculating only the parts contributing to the final bitstrings.

We observe that the order of dimensions significantly affects the computation time of cuTensor. Empirically, better performance is achieved when satisfying the following conditions: 1. Placing the contracted dimensions of input tensors at the end. 2. Ensuring the same order of the contracted dimensions between two input tensors. 3. Arranging the order of dimensions of input tensors according to the order of dimensions of output tensors. We employ a greedy algorithm to optimize the order of dimensions of all tensors within the contraction tree, starting from the root node. 
Additionally, we used cuTensor's einsum~\cite{pan2023efficient} in the contraction process. For each contraction step, we compared the performance of transpose-transpose-GEMM-transpose (TTGT) and cuTensor and chose to use the faster of the two.
% we compare the performance of cuTensor with transpose-transpose-GEMM-transpose (TTGT) for steps where TTGT exhibits significantly faster performance and substitute TTGT when applicable.

Our algorithm utilizes 8 GPUs on a single node for computation, with each node handling and accumulating different sliced subtasks. To obtain the final result, we adopt NVIDIA Collective Communication Library (NCCL) reduction based on the ring-reduce method. Its communication complexity is $2(N-1)\times K/N$, where $K$ represents the data volume per node, and $N$ is the number of nodes. Therefore, as the number of nodes increases, the communication time tends toward a constant value, which in our tests is approximately 2 seconds. Hence, reduction time does not become a main bottleneck.

For the case of an 80 GB memory constraint with a single A100 GPU, we find the optimal contraction scheme comprises $2^{30}$ subtasks, with each subtask taking 3.95 seconds to compute. With the constraint of 640GB memory and 8 A100 GPUs, the optimal scheme consists of $2^{24}$ subtasks, and each subtask takes 2.95 seconds to compute on a single node (8 A100 GPUs). Therefore, we estimate that our optimized 8-GPU parallel algorithm yields a speedup of  $3.95 \times 2^6 / (2.95 \times 8) = 10.7$ times.

\subsection{Parallelism and complexity for simulating
the \textit{Sycamore}}
\begin{figure}[htb]
\centering
\includegraphics[width=0.45\textwidth]{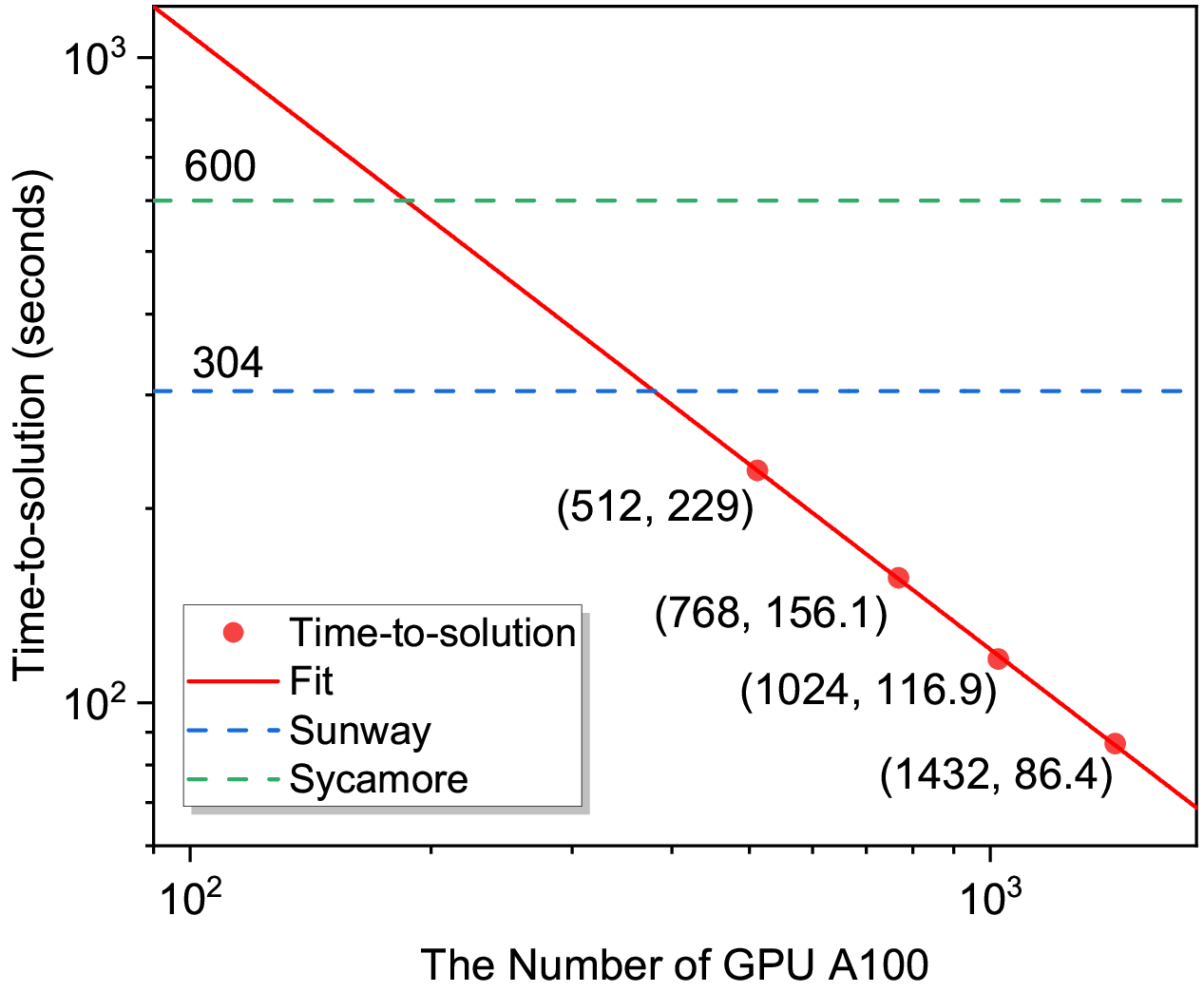}% Here is how to import EPS art
\caption{\label{linearl_decrease_time} 
The time-to-solution for simulating a quantum RCS experiment with $XEB = 0.002$ for samples obtained via the post-processing method on/using different numbers of GPUs. The time-to-solution decreases linearly as the number of GPUs increases, which demonstrates the parallelism of our algorithm.
}
\end{figure}

To sample from the \textit{Sycamore} circuit, we choose 10 qubits as open qubits to generate $\tilde{\mathcal{S}}$ with $|\tilde{\mathcal{S}}|=(3\cdot10^{6})\times 2^{10}$ and aim to obtain $k=3\cdot 10^{6}$ uncorrelated samples via the top-$k$ method. 
As mentioned above, the large tensor network could be split into independent sub-networks and each sub-network can be contracted on a single GPU. 
In our implementation, thanks to the approximate simulation and the post-processing method, we only need to perform $0.03\%$ of the total tasks (i.e., contract $0.03\%$ sub-networks of total networks) out of a total of $2^{24}$ subtasks to reach an XEB value of 0.002, where each contraction subtask requires $1.1029\times10^{14}$ FLOPS and can be completed in 3.09 seconds on 8 NVIDIA A100 GPUs.
% Each contracting subtask exhibits a computational complexity of $1.1029\times10^{14}$ FLOPS. Since the approximate simulation and the post-processing algorithm, we only need to select 5012 subtasks from the total $2^{24}$ subtask to calculate. On 8 NVIDIA A100 GPUs, the computation of a single subtask requires 3.09 seconds.
Furthermore, we observe a linear decay of the total time-to-solution as we increase the number of GPUs used in computation (Fig.~\ref{linearl_decrease_time}), which clearly demonstrates the parallelism of the implementation of our algorithm. In particular, when using 1,432 GPUs, the total time-to-solution for sampling $3\times10^6$ bitstrings is 86.4 sec, which is less than \textit{Sycamore}'s 600-second time-to-solution.

To verify the fidelity, 
we utilize the exact output amplitudes from a previous study~\cite{liu2022validating} to calculate the fidelity of our simulation of the \textit{Sycamore} circuit with 53 qubits and 20 layers of gates, and we observe that the numerical fidelity $F_{\text{num}} = 0.01823\%$ fits well with our prediction $F_{\text{predicted}}=0.01907\%$ according to Eq.~\ref{eq: XEB top-k}. 
% we utilize the exact amplitudes from a previous study~\cite{liu2022validating} to calculate the numerical fidelity of the simulation of the 53 qubits 20 layers of gate circuit. We observe that when the predicted fidelity $F_{\text{predicted}}=0.01907\%$, the corresponding numerical fidelity $F_{\text{num}} = 0.01823\%$.

% \subsection{Fidelity of approximate simulation}
% The fidelity of the numerical results is highly consistent with the estimated fidelity $F_{\text{estimated}}$, which corresponds to the fraction $f$ of the tensor network taken into consideration $F_{\text{estimated}}=f$. We have researched some different cases with $f=1/2^D$. The consistency implies that the tensor network can be divided into distinct parts, with each part contributing equally to the fidelity. Consequently, the total fidelity is determined by the number of contraction parts ~\cite{Pan2022PhysRevLett.129.090502}.

The complexity of this task is summarized in Tab. \ref{complexity}. The space complexity of our algorithm is proportional to $s2^M$, where $s$ represents the size of the data type and $M$ corresponds to the targeted contraction treewidth, which can be specified manually in our slicing algorithm. So the space complexity is completely under control.
% Furthermore, the time complexity is characterized by $O(fMm2^{n}/ln(\alpha))$, where $m$ denotes the number of gate layers, $n$ represents the number of qubits, $f$ represents the fidelity, and $\alpha$ denotes the post-processing coefficient. 
% One notable advantage of our algorithm is the reduction in time complexity through approximate simulation and post-processing algorithms. 
% Comparing with this work~\cite{Pan2022PhysRevLett.129.090502}, time complexity of our algorithm decreases $\ln (|\tilde{\mathcal{S}}|/k)$ by post-processing method.
% Additionally, our algorithm allows for complete control over the space complexity. 

\section{CONCLUSIONS}\label{sec13}

We reduced the computational complexity by 7 times using the post-selection algorithm. Because the post-selection algorithm can filter out high XEB samples from low-fidelity pre-samples, while low-fidelity sampling only has low complexity.
In addition, according to the observation that the compute complexity of the optimal contraction scheme is inversely proportional to the storage space, 8-GPU interconnection technology is developed to achieve 640G memory space. We took advantage of the large storage space to increase the speed of parallel algorithms by 10.7 times.
To finish the same task as \textit{Sycamore}, we run our algorithm on 1,432 NVIDIA A100 GPUs and obtain 3 million uncorrelated samples with XEB values of $2\times 10^{-3}$ in 86.4 seconds with an energy consumption of 13.7 kWh. In this sense, we estimate that approximately 206 GPUs have an equivalent computational power in implementing a 53-qubit 20-depth random circuit sampling as \textit{Sycamore} (Fig. \ref{linearl_decrease_time}). With further optimizations, the power consumption is expected to be reduced below that of Sycamore. Our results suggest that establishing quantum advantage requires larger-scale quantum experiments.

{ACKNOWLEDGMENTS}

We wish to thank Yao-Jian Chen, and Yu-Xuan Li for their valuable discussions.

Our work is supported by the National Natural Science Foundation of China, the National Key R\&D Program of China, the Chinese Academy of Sciences, the Anhui Initiative in Quantum Information Technologies, the Science and Technology Commission of Shanghai Municipality, the Innovation Program for Quantum Science and Technology (Grant No. 2021ZD0301400).

\begin{table*}[bht]
  \begin{center}
    \caption{The complexity of simulating quantum RCS experiment. The task with $XEB_{p.p.}=0.002$ corresponds to the simulation of \textit{Sycamore}\cite{Arute2019}. The efficiency is calculated by $8*T_c/(P\cdot t)$, where 8 is the complex number product factor, $T_c$ is the time complexity, P is the single-precision performance of GPU and t is the time-to-solution.\label{complexity}}
    \begin{tabular}{|c|c|c|c|}
        \hline & each sub-network & $\begin{array}{c}\textbf{\textit{Sycamore}}\\(XEB_{p.p.}=0.002) \end{array}$ & Fidelity $=1$ \\
        \hline $\begin{array}{c}\text { Time complexity } \\
        \text { (FLOP) }\end{array}$ & $ 1.1029 \times 10^{14} $ & $ 5.5527\times 10^{17} $ & $ 1.8504\times 10^{21} $ \\
        \hline $\begin{array}{c}\text { Space complexity } \\
        \text { (GB) }\end{array}$ & \multicolumn{3}{|c|}{640} \\
        \hline $\begin{array}{c}\text { Memory complexity } \\
        \text { (Bytes) }\end{array}$ & $ 1.5200\times 10^{12} $ & $ 7.6182\times 10^{15} $ & $ 2.5668\times 10^{19} $ \\
        \hline Efficiency & $23.57\%$ & $22.12\%$ & - \\
        \hline $\begin{array}{c}\text { Time-to-solution } \\
        \text { (sec) }\end{array}$ & 2.95 & 86.4 & - \\
        \hline $\begin{array}{c}\text { Computer resource } \\
        \text { (A100) }\end{array}$ & 8 & 1,432 & - \\
        \hline
    \end{tabular}
  \end{center}
\end{table*}

\bibliography{apssamp}% Produces the bibliography via BibTeX.

\appendix    % this command starts appendixes

\section{The theory value of XEB after post-processing}
\label{appendix: The theory value of XEB
after post-processing}
We assume that the bitstring probability $q$ follows the Porter-Thomas distribution:
$$
P(Nq)=e^{-N q}
$$
with $N=2^{n}$ and $n$ is the number of qubits. Let us set $x=N q$, then we have $P(x)=e^{-x}$.
we select k bitstrings with the top k probabilities from $N$ bitstrings, we can determine the minimum probability $t/N$ among the selected bitstrings as
$$
\int_t^{\infty} e^{-x} d x=k/N
$$
which gives $t=\ln N/k$, we set $k/N=\alpha $.
It is easy to check that the expectation of the $\alpha$ fraction of bitstrings is
$$
\int_t^{\infty} x e^{-x} d x=\alpha(1-\ln \alpha)
$$
Then we can calculate XEB of the $\alpha$-fraction bistrings
$$
\begin{aligned}
 \mathcal{F}_{\text{XEB,top-k}}^{\textit{ideal}} & =\frac{N}{\alpha N} \sum_{i=1}^{\alpha N} q_i-1 \\
& =\frac{N}{\alpha N} \int_t^{\infty} x e^{-x} d x-1 \\
& =-\ln \alpha
\end{aligned}
$$
where $i=1,2, \cdots, \alpha N$ denotes indices of top $\alpha N$ bitstrings.
If the state is an approximation with fidelity $F$, then we expect that the XEB is
$$
\begin{aligned}
    \mathcal{F}^{\textit{approx}}_{\text{XEB,top-k}}
    &=-F \ln \alpha\\
    &= F\ln N/k
\end{aligned}
$$
so we have proven the equation \ref{eq: XEB top-k}.
\end{document}